\documentclass[prd,amsmath,amssymb,reprint,preprintnumbers,nofootinbib,superscriptaddress]{revtex4-1}
              
\pdfoutput=1

\usepackage{amsmath}
\usepackage{amsfonts}
\usepackage{amssymb}
\usepackage{mathrsfs}
\usepackage{graphicx}
\usepackage{color}
\usepackage[dvipsnames]{xcolor}
\usepackage{longtable}
\usepackage{bm}
\usepackage{blindtext}
\usepackage{wasysym}
\usepackage{hyperref}
\hypersetup{colorlinks=true,allcolors=blue}
\usepackage[normalem]{ulem}

\newcommand\Fontx{\fontsize{10}{12}\selectfont}

\newcommand{\ie}{\textit{i.e.}}

\begin{document}
 
\title{Flavor-Dependent Long-Range Neutrino Interactions in DUNE and T2HK: \\ Synergy Breeds Power\\\vskip0.2cm
\Fontx{``Contribution to the 25th International Workshop on Neutrinos from Accelerators''}}
\author{Masoom Singh}
\email{masoom@iopb.res.in (ORCID: 0000-0002-8363-7693)}
\affiliation{Department of Physics, Utkal University, Vani Vihar, Bhubaneswar 751004, India}
\affiliation{Institute of Physics, Sachivalaya Marg, Sainik School Post, Bhubaneswar 751005, India}
%%%%
\author{Mauricio Bustamante}
%\email{anil.kumar@desy.de (ORCID: 0000-0002-8367-8401)}
\affiliation{Niels Bohr International Academy, Niels Bohr Institute, University of Copenhagen, DK-2100 Copenhagen, Denmark}

%%%%
\author{Sanjib Kumar Agarwalla}
%\email{sanjib@iopb.res.in (ORCID: 0000-0002-9714-8866)}
\affiliation{Institute of Physics, Sachivalaya Marg, Sainik School Post, Bhubaneswar 751005, India}
\affiliation{Homi Bhabha National Institute, Anushakti Nagar, Mumbai 400094, India}
\affiliation{Department of Physics \& Wisconsin IceCube Particle Astrophysics Center, University of Wisconsin, Madison, WI 53706, U.S.A}
%%%%

\preprint{NuFact 2024-29}

\date{\today}

\begin{abstract}
	
\begin{center} (presented by Masoom Singh)\end{center}

Discovering new neutrino interactions would provide evidence of physics beyond the Standard Model. We focus on flavor-dependent long-range neutrino interactions mediated by ultra-light mediators (masses below $10^{-10}$~eV) from lepton-number gauge symmetries $L_e-L_\mu$, $L_e-L_\tau$, and $L_\mu-L_\tau$. These interactions, sourced by electrons and neutrons in the Earth, Moon, Sun, Milky Way, and the local Universe, could modify neutrino oscillation probabilities. The upcoming long-baseline experiments, DUNE and T2HK, with their large statistics, reduced systematic uncertainties, and well-characterized neutrino beams, will probe these interactions. We forecast that, while individually DUNE and T2HK could constrain these long-range neutrino interactions, their combination lifts parameter degeneracies that weaken individual sensitivity and provides stronger constraints.
  
\end{abstract}

\maketitle

%%%%%%%%%%%%%%%%%%%%%%%%%%%%%%%%%%%%%%%%%%%%%

%%%%%%%%%%%%%%%%%%%%%%%%%%%%%%%%%%%%%%%%%%%%%
\section{Introduction} 
%%%%%%%%%%%%%%%%%%%%%%%%%%%%%%%%%%%%%%%%%%%%%

Uncovering the mysteries of flavor-dependent neutrino-matter interactions offers a promising pathway to explore physics that extends beyond the Standard Model (SM). Such interactions influence neutrino flavors $\left(\nu_{e},\, \nu_{\mu}, \, \nu_{\tau}\right)$ in distinct ways. These variations impact the effective Hamiltonian governing neutrino propagation, ultimately altering the probabilities of neutrino oscillations $-$ a phenomenon that can provide vital clues to new physics. 

A unique way to generate these flavor-dependent interactions involves gauging specific anomaly-free global $U$(1) symmetries in the SM. These symmetries emerge from combinations of lepton numbers $\left(L_{e},\, L_{\mu},\, L_{\tau}\right)$ and baryon number $(B)$. Introducing a new neutral vector boson, $Z^\prime$, these interactions become long-range when the mass of $Z^\prime$ is extraordinarily small, in the range $[10^{-35}, 10^{-10}]$ eV. At such tiny masses, the interaction spans vast distances, allowing contributions to the neutrino-matter potential from both local and distant sources, ranging from nearby celestial bodies like the Earth and Moon, to more distant objects such as the Sun, the Milky Way, and even the cosmological matter distribution. Detecting these interactions, despite their faintness, becomes feasible in experiments capable of observing minuscule deviations in neutrino oscillation probabilities. Long-baseline (LBL) neutrino oscillation experiments, including the upcoming DUNE and T2HK, are particularly well-suited for this task. These next-generation experiments combine high-precision detection capabilities with well-characterized neutrino beams, enabling the exploration of sub-leading effects that were previously beyond the reach of existing facilities.

Our analysis emphasizes the significance of these interactions in the context of  $U(1)$ symmetries such as $L_{e}-L_{\mu},\, L_{e}-L_{\tau},\,$ and $L_{\mu}-L_{\tau}$. These symmetries, stemming from fundamental extensions of the SM, introduce flavor-dependent effects primarily on neutrino interactions with electrons or neutrons.

%%%%%%%%%%%%%%%%%%%%%%%%%%%%%%%%%%%%%%%%%%%%%
\section{Flavor-dependent long-Range neutrino Interactions} 
%%%%%%%%%%%%%%%%%%%%%%%%%%%%%%%%%%%%%%%%%%%%%

In the SM, the baryon ($B$) and lepton numbers ($L_e$, $L_\mu$, and $L_\tau$) are accidental $U(1)$ global symmetries. Linear combinations of these lepton-number symmetries can be gauged in an anomaly-free way, allowing for new interactions without introducing new fermions or right-handed neutrinos, though not simultaneously. The DUNE and T2HK experiments provide an opportunity to explore such anomaly-free $U(1)$ gauge symmetries, including $L_e-L_\mu$, $L_e-L_\tau$, and $L_\mu-L_\tau$. These symmetries introduce flavor-dependent neutrino-matter interactions, affecting neutrino oscillations.

These interactions can be expressed in the form of flavor-dependent Yukawa potentials, sourced by electrons and neutrons, which influence neutrino flavor transition. For the $L_e-L_\beta$ interaction, where $\beta = \mu$ or  $\tau$, the neutrino experiences a potential:
\begin{equation}
 V_{e\beta}
 =
 G_{\alpha\beta}^{\prime 2}
 \frac{N_e}{4\pi d}
 e^{-m'_{e\beta}d} \;,
 % \left(4\pi d\right)^{-1}e^{-m'_{e\beta}d}\,,
 \label{equ:potential_e_beta}
\end{equation}
where $d$ is the distance of neutrino from $N_{e}$ electron density, $m_{e\beta}^\prime$ is the mass of the mediating $Z_{e\beta}^\prime$ boson.  Similarly, for the $L_\mu-L_\tau$\, interaction, the potential is
\begin{equation}
 V_{\mu\tau}
 =
 G_{\alpha\beta}^{\prime 2}
  \frac{e}{\sin\theta_W\cos\theta_W}
 \frac{N_n}{4 \pi d}
 e^{-m'_{\mu\tau}d} \;,
 \label{equ:potential_mu_tau}
\end{equation}
where $m_{\mu\tau}^\prime$ is the mass of the mediating $Z_{\mu\tau}^\prime$ boson.

In the above equations, the effective coupling strength, $G^\prime_{\alpha\beta}$, for different neutrino flavors can be expressed as: \begin{equation}
 G_{\alpha \beta}^\prime
 =
 \left\{
  \begin{array}{lll}
   g^{\prime }_{e \mu} & , & ~{\rm for}~\alpha = e, \beta = \mu \\
   g^{\prime }_{e \tau} & , & ~{\rm for}~\alpha = e, \beta = \tau \\
   \sqrt{g^{\prime}_{\mu \tau} (\xi-\sin \theta_W \chi)} & , & ~{\rm for}~\alpha = \mu, \beta = \tau \\
  \end{array}
 \right. \;.
 \label{equ:Gab}
\end{equation}

The terrestrial neutrinos can feel the potential due to the Earth, Moon, Sun, Milky Way, and local Universe depending on the interaction range, and the total potential is the sum of contribution from all these sources:

\begin{equation}
 \label{equ:pot_total}
 V_{\alpha \beta}
 =
 V_{\alpha \beta}^\oplus + V_{\alpha \beta}^{\leftmoon} + V_{\alpha \beta}^{\astrosun} + V_{\alpha \beta}^{\rm MW} +  V_{\alpha \beta}^{\rm cos} \;,
\end{equation}
where $V_{\alpha \beta}^\oplus, \, V_{\alpha \beta}^{\leftmoon},$ and $V_{\alpha \beta}^{\astrosun}$ represent the potentials sourced by the Earth, Moon, and Sun, respectively, and $V_{\alpha \beta}^{\rm MW}$ and $ V_{\alpha \beta}^{\rm cos}$ are from the Milky Way and the cosmological matter distribution. We compute the average potential at the point of detection, especially for light mediators which correspond to interaction ranges from hundreds of meters to the observable Universe. The potential sources are treated as electrically neutral and isoscalar, where the number of electrons and protons, as well as neutrons, are assumed to be equal, except for the Sun and cosmological matter. The Hamiltonian driving neutrino propagation, including both the SM and new long-range interactions, is:
\begin{equation}
 \label{equ:hamiltonian_tot}
 \mathbf{H}
 =
 \mathbf{H}_{\rm vac}
 +
 \mathbf{V}_{\rm mat}
 +
 \mathbf{V}_{\alpha\beta} \;,
\end{equation}
where $\mathbf{H}_{\rm vac}$ is in vacuum and $\mathbf{V}_{\rm mat}$ depicts contribution from the SM matter effects. The contribution from the new interaction is:
\begin{equation}
 \label{equ:pot_lri_matrix}
 \mathbf{V}_{\alpha\beta}
 =
 \left\{
  \begin{array}{ll}
   {\rm diag}(V_{e\mu}, -V_{e\mu}, 0), & {\rm for}~ \alpha, \beta = e, \mu \\
   {\rm diag}(V_{e\tau}, 0, -V_{e\tau}), & {\rm for}~ \alpha, \beta = e, \tau \\
   {\rm diag}(0, V_{\mu\tau}, -V_{\mu\tau}), & {\rm for}~ \alpha, \beta = \mu, \tau \\   
  \end{array}
 \right. \;,
\end{equation}
For antineutrinos, it flips sign, \ie, $\mathbf{V}_{\alpha\beta} \to -\mathbf{V}_{\alpha\beta}$. For further details, see Ref.~\cite{Singh:2023nek}

%%%%%%%%%%%%%%%%%%%%%%%%%%%%%%%%%%%%%%%%%%%%%%%%%%%%%%%%%%%%%%%%%%%%%%%%%%%%%%%
\section{Results}
%%%%%%%%%%%%%%%%%%%%%%%%%%%%%%%%%%%%%%%%%%%%%%%%%%%%%%%%%%%%%%%%%%%%%%%%%%%%%%%

We evaluate the sensitivity of the DUNE and T2HK experiments to detect the presence of these long-range neutrino interactions, which manifest through alterations in neutrino oscillation probabilities and their corresponding rate of neutrino detection. By assuming the absence of long-range interactions, we establish constraints on the associated matter potential in terms of the mediator mass and coupling strength. Additionally, we analyze the impact of parameter degeneracies between the long-range interaction potential and the CP-violating phase, $\delta_{\rm CP}$\,, and atmospheric mixing angle, $\sin^2\theta_{23}$\,, in quantifying the discovery potential of these novel interactions.

%%%%%%%%%%%%%%%%%%%%%%%%%%%%%%%%%%%%%%%%%%%%
\begin{figure}
\includegraphics[width=\linewidth]{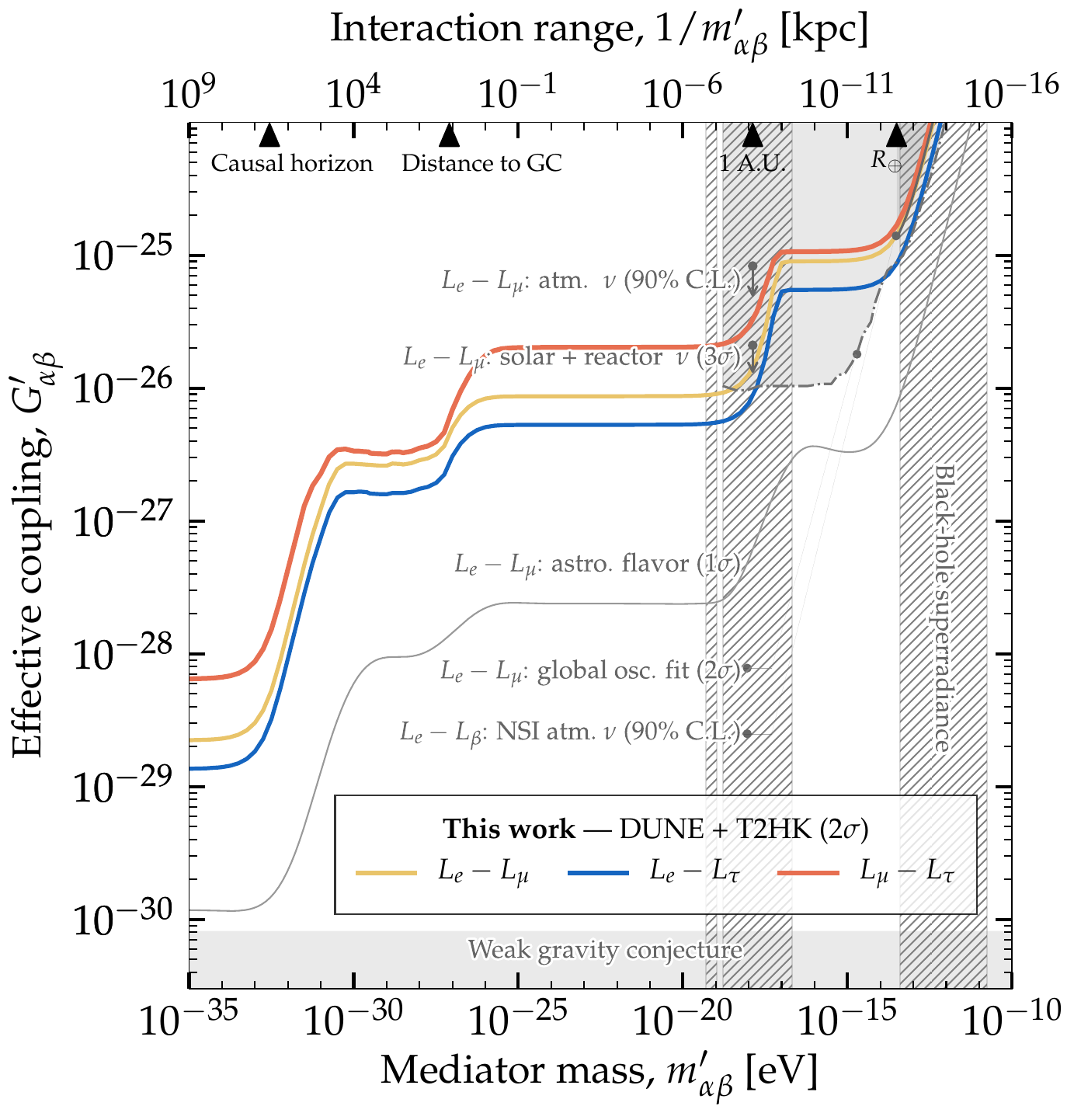}
\caption{We depict the projected upper bounds on the effective coupling $G_{\alpha\beta}^{\prime}$ as a function of mediator mass $m_{\alpha\beta}^\prime$ associated with the new neutral mediator boson, $Z_{\alpha\beta}^\prime$. These projections are derived from the combined analysis of the simulated data from the DUNE and T2HK experiments. The DUNE configuration assumes an operational period of 5 years each for $\nu$ and $\bar{\nu}$ modes, while T2HK operates for 2.5 years in $\nu$ mode and 7.5 years in $\bar{\nu}$ mode. For this analysis, we generate the prospective data assuming $\delta_{\rm CP} = 223^{\circ}$, $\sin^2\theta_{23} = 0.455$, and normal neutrino mass ordering. Existing constraints on $G_{\alpha\beta}^\prime$ from global neutrino oscillation fits~\cite{Coloma:2020gfv}, atmospheric~\cite{Joshipura:2003jh}, solar and reactor data~\cite{Bandyopadhyay:2006uh}, non-standard interactions~\cite{Super-Kamiokande:2011dam, Ohlsson:2012kf, Gonzalez-Garcia:2013usa}, projected sensitivity from IceCube-Gen2~\cite{Bustamante:2018mzu,Agarwalla:2023sng}, and indirect limits from black-hole superradiance~\cite{Baryakhtar:2017ngi} and weak gravity conjecture~\cite{Arkani-Hamed:2006emk} are also shown. For detailed explanation see the elaborated text in Ref.~\cite{Singh:2023nek}. }
\label{Fig:1}
\end{figure}
%%%%%%%%%%%%%%%%%%%%%%%%%%%%%%%%%%%%%%%%%%%%%%

%%%%%%%%%%%%%%%%%%%%%%%%%%%%%%%%%%%%%%%%%%%%
\begin{figure*}[htb!]
\includegraphics[width=0.9\textwidth]{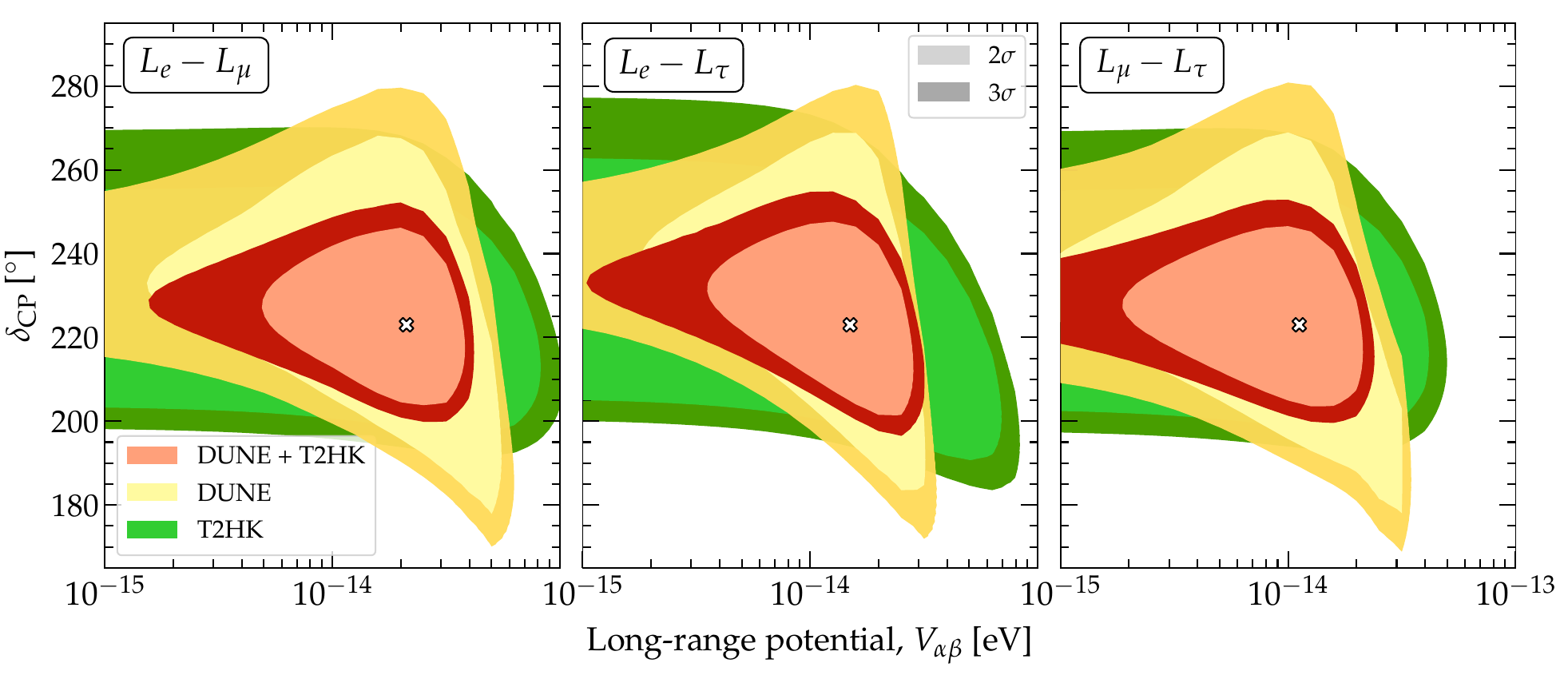}
\caption{We project the allowed regions of the long-range potential, $V_{\alpha\beta}$ and the CP phase, $\delta_{\rm CP}$, assuming a fixed true value of $V_{\alpha\beta}$ for each symmetry and $\delta_{\rm CP} (\rm true) = 223^{\circ}$. Results are shown at both 2$\sigma$ and 3$\sigma$ confidence levels. The combination of DUNE and T2HK highlights their complementarity, enabling robust constraints across the planes of $V_{\alpha\beta}$ - $\delta_{\rm CP}$ for all the three symmetries. This figure is taken from Ref.~\cite{Singh:2023nek}.}
\label{Fig:2}
\end{figure*}
%%%%%%%%%%%%%%%%%%%%%%%%%%%%%%%%%%%%%%%%%%%%

Figure~\ref{Fig:1} presents the translated 2$\sigma$ upper bounds on $V_{\alpha\beta}$ on the plane of the effective coupling $G_{\alpha\beta}^{\prime}$ and mediator mass plane $m_{\alpha\beta}^\prime$. The yellow, blue, and red colored curves represent the isocontours of new interaction potential $V_{e\mu} = 1.4 \times 10^{-14}$ eV, $V_{e\tau} = 1 \times 10^{-14}$~eV, and $V_{\mu\tau} = 0.73 \times 10^{-14}$ eV, which correspond to the 2$\sigma$ upper bounds on  $V_{\alpha\beta}$, using DUNE + T2HK, respectively. 

The step-like transitions observed across different values of $m_{\alpha\beta}^\prime$ correspond to the extension of the interaction range, allowing it to encompass additional sources of electrons or neutrons. This extended range effectively introduces new contributions to the interaction potential, $V_{\alpha\beta}$ thereby altering its overall strength and profile. Below $m_{\alpha\beta}^\prime\sim 10^{-18}$ eV, our projected limits probe a parameter space that remains largely uncharted by past terrestrial experiments. The constraints projected from DUNE and T2HK demonstrate exceptional sensitivity, due to their large event rates, controlled systematics, and well-characterized neutrino beams. Although IceCube-Gen2 is expected to have superior sensitivity due to high-energy astrophysical neutrinos in the TeV-PeV range, presently this potential is mitigated by significant astrophysical uncertainties and limited event rates. These challenges are likely to be overcome in the near-future. 

We find that the limits on $V_{e\tau}$ for the $L_{e}-L_{\tau}$ symmetry are stringent as these are mainly driven by the $\nu_{\mu}\rightarrow \nu_{e}$ and $\bar{\nu}_{\mu} \rightarrow \bar{\nu}_{e}$ appearance probabilities. For the $L_{e}-L_{\mu}$ symmetry, although the associated interaction affects both appearance and disappearance probabilities, but it has a smaller impact compared to the $L_{e}-L_{\tau}$ symmetry. As a result, the constraints on $V_{e\mu}$ are weaker than $V_{e\tau}$. Note that the presence of $V_{\mu\tau}$ potential arising from $L_{\mu}-L_{\tau}$ symmetry only influence the $\nu_{\mu}\rightarrow \nu_{\mu}$ and $\bar{\nu}_{\mu} \rightarrow \bar{\nu}_{\mu}$ disappearance probabilities. Unlike the other two symmetries, the $G^\prime_{\mu\tau}$ depends on $\sqrt{g^{\prime}_{\mu \tau} (\xi-\sin \theta_W \chi)}$ (see Eq.~\ref{equ:Gab}). Thus scanning over it as a whole provides the weakest constraints. 

%%%%%%%%%%%%%%%%%%%%%%%%%%%%%%%%%%%%%%%%%%%%
\begin{figure*}[ht!]
\includegraphics[width=0.87\textwidth]{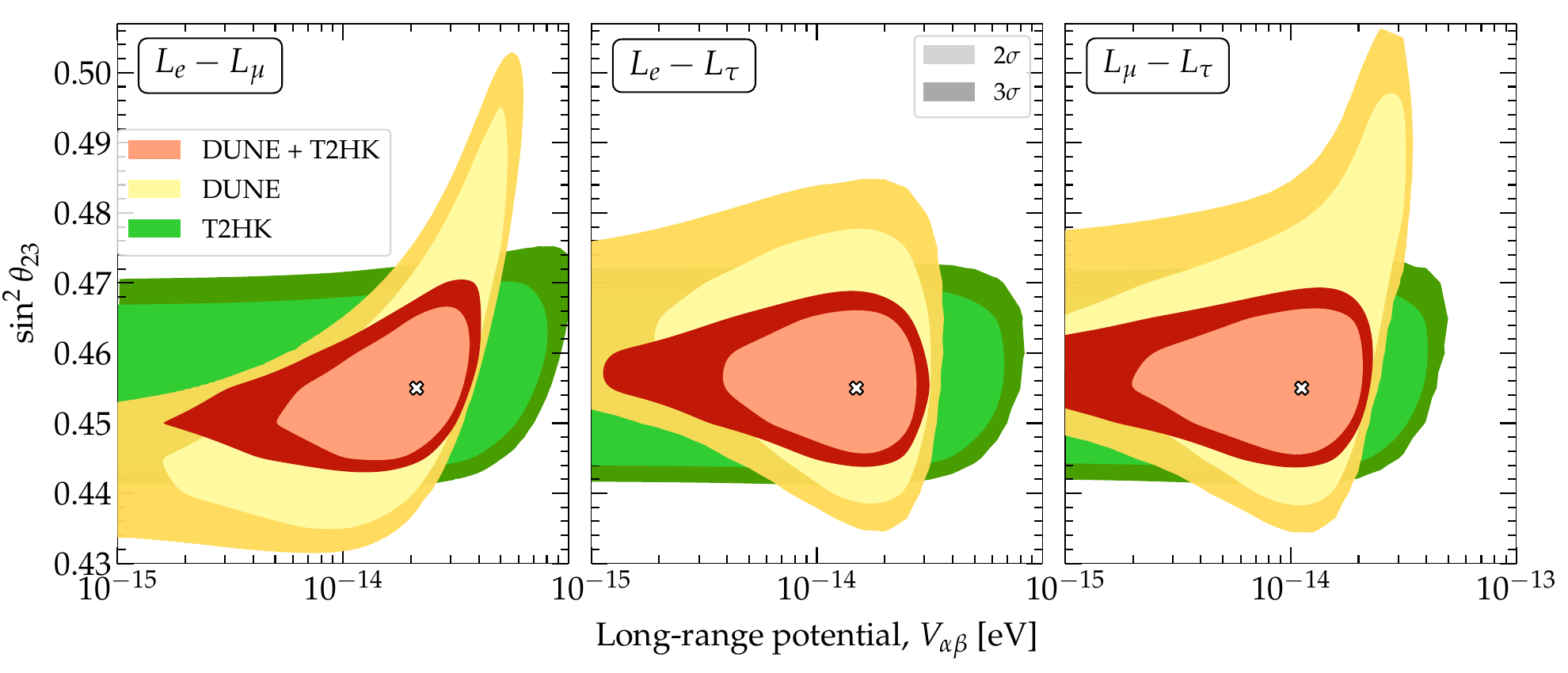}
\caption{We project the allowed regions of the long-range potential, $V_{\alpha\beta}$ and  $\sin^2\theta_{23}$, assuming a fixed true value of $V_{\alpha\beta}$ for each symmetry and $\sin^2\theta_{23} (\rm true)= 0.455$. Results are shown at both 2$\sigma$ and 3$\sigma$ confidence levels. The combination of DUNE and T2HK highlights their complementarity, enabling robust constraints across the planes of $V_{\alpha\beta}$ - $\sin^2\theta_{23}$ for all the three symmetries. This figure is taken from Ref.~\cite{Singh:2023nek}.
}
\label{Fig:3}
\end{figure*}
%%%%%%%%%%%%%%%%%%%%%%%%%%%%%%%%%%%%%%%%%%%%

Figures~\ref{Fig:2} and~\ref{Fig:3} represent the allowed regions in $\left(V_{\alpha\beta}-\delta_{\rm CP}\right)$ and $\left( V_{\alpha\beta}-\sin^2\theta_{23}\right)$ planes, respectively. We generate the prospective data by illustratively fixing the true value of the long-range potentials: $V_{e\mu}^{\rm true}$ = $2.12 \times 10^{-14}$~eV (left panel), $V_{e\tau}^{\rm true}$ = $1.6 \times 10^{-14}$~eV (middle panel), and $V_{\mu\tau}^{\rm true}$ = $1.12 \times 10^{-14}$~eV (right panel). Also, the true choice of $\delta_{\rm CP} = 223^\circ$ in Fig.~\ref{Fig:2} and the true value of $\sin^{2}\theta_{23} = 0.455$ in Fig.~\ref{Fig:3} following Ref.~\cite{Capozzi:2021fjo}. The true $V_{\alpha\beta}$ values correspond to the 3$\sigma$ upper limits achievable by DUNE + T2HK in constraining the long-range potential under each symmetry~\cite{Singh:2023nek}. While determining the sensitivity, we vary $V_{\alpha\beta}$ in the range $[3\times 10^{-14} , 10^{-15}]$ eV, where its impact is potentially visible in neutrino flavor transitions as observed by DUNE and T2HK. We minimize over the 3$\sigma$ uncertainties of $|\Delta m^2_{31}| \in [2.438, 2.602] \times 10^{-3}$~eV$^2$ and both the choices of mass ordering in the test-statistic of Figs.~\ref{Fig:2} and~\ref{Fig:3}. Additionally, in Fig.~\ref{Fig:2}, we minimize over the 3$\sigma$ uncertainties of $\sin^2\theta_{23} \in [0.4, 0.6]$, and in Fig.~\ref{Fig:3}, over $\delta_{\rm CP} \in [139^\circ, 355^\circ]$.

The degeneracies among the uncertainties in $\delta_{\rm CP}$, $\theta_{23}$ and the neutrino mass ordering with the new long-range potential, $V_{\alpha\beta}$\,, are responsible for deteriorating the sensitivity while constraining the parameter values. Figures~\ref{Fig:2} and~\ref{Fig:3} depict that DUNE exhibits remarkable sensitivity in constraining $V_{\alpha\beta}$, owing to its broad energy coverage and large matter effect. On the other hand, T2HK provides unparalleled precision in constraining the CP-violating phase $\delta_{\rm CP}$ and $\sin^2\theta_{23}$ driven by its large event statistics and enhanced sensitivity to both appearance and disappearance channels. The smaller systematic uncertainties of DUNE in appearance channel and T2HK in disappearance channel also play a crucial role.

 The combination of DUNE and T2HK demonstrates their complementary strengths~\cite{Agarwalla:2022xdo}, enabling comprehensive and robust constraints across both the $V_{\alpha\beta}$-mediated interaction and $\delta_{\rm CP}$ and $\sin^2\theta_{23}$ parameter spaces.

%%%%%%%%%%%%%%%%%%%%%%%%%%%%%%%%%%%%%%%%%%%%%%%%%%%%%%%%%%%%%%%%%%%%%%%%%%%%%%%
\section{Conclusions}
%%%%%%%%%%%%%%%%%%%%%%%%%%%%%%%%%%%%%%%%%%%%%%%%%%%%%%%%%%%%%%%%%%%%%%%

Neutrinos, could provide crucial insights into physics beyond the Standard Model, especially through potential interactions beyond the weak force. Next-generation long-baseline experiments like DUNE and T2HK offer a promising opportunity to explore new flavor-dependent neutrino interactions, particularly those introduced by gauging $L_e-L_\mu$, $L_e-L_\tau$, and $L_\mu-L_\tau$ $U(1)$ symmetries, which are anomaly-free and mediated by a neutral vector boson. These long-range interactions, involving ultra-light mediators with masses below $10^{-10}$~eV, could subtly affect neutrino oscillations, making them testable in future experiments. Our forecasts highlight the complementary capabilities of DUNE and T2HK, showing that while each experiment has limitations due to parameter degeneracies, their combination could provide a more robust probe, offering the potential to discover long-range neutrino interactions and setting tighter constraints on their properties. 

\begin{acknowledgments}
We acknowledge financial support from DAE, DST,
DST-SERB, Govt. of India, INSA, and USIEF.
M.B. is supported by the {\sc Villum Fonden} under the project no. 29388. The numerical simulations are performed using the “SAMKHYA: High-Performance Computing Facility” provided by the Institute of Physics, Bhubaneswar, India.
\end{acknowledgments}
%%%%%%%%%%%%%%%%%%%%%%%%%%%%%%%%%%%%%%%%%%%%%%%%%%%%%%%%%%%%%%%%%%%%%%%%%%%%%%%

\bibliographystyle{apsrev4-1}
\bibliography{LRI-LBL-three-symmetries-NuFACT2024.bib}

\end{document}